# The Role of $^{11}$B$_4$C Interlayers in Enhancing Fe/Si Multilayer Performance for Polarized Neutron Mirrors

A. Zubayer[1], F. Eriksson[1], M. Falk[1], M. Lorentzon[1], J. Palisaitis[1], C. Klauser[2], G. Nagy[4], P. M. Wolf[4], E. Pitthan[4], R. Holeňák[4], D. Primetzhofer[4], G.B.G. Stenning[3], A. Glavic[2], J. Stahn[2], S. Dorri[1], P. Eklund[1,4], J. Birch[1], N. Ghafoor[1]

1. Thin Film Physics Division, Department of Physics, Chemistry, and Biology (IFM), Linköping University, SE-581 83 Linköping, Sweden
2. PSI Center for Neutron and Muon Sciences, 5232 Villigen PSI, Switzerland
3. ISIS Neutron and Muon Source, Rutherford Appleton Laboratory, Didcot OX11 0QX, United Kingdom
4. Ångström Laboratory, Uppsala University, Box 538, SE-751 21 Uppsala, Sweden

## Abstract

This study investigates the effects of incorporating $^{11}$B$_4$C interlayers into Fe/Si multilayers, with a focus on interface quality, reflectivity, polarization, and magnetic properties for polarized neutron optics. It is found that the introduction of 1 Å and 2 Å $^{11}$B$_4$C interlayers significantly improves the interface sharpness, reducing interface width and preventing excessive Si diffusion into the Fe layers. X-ray reflectivity and polarized neutron reflectivity measurements show enhanced reflectivity and polarization, with a notable increase in polarization for 30 Å period multilayers. The inclusion of interlayers also helps prevent the formation of iron-silicides, improving both the magnetic properties and neutron optical performance. However, the impact of interlayers is less pronounced in thicker-period multilayers (100 Å), primarily due to the ratio between layer and interface widths. These results suggest that $^{11}$B$_4$C interlayers offer a promising route for optimizing Fe/Si multilayer performance in polarized neutron mirrors.

## 1. Introduction

Neutron scattering is a powerful technique for studying atomic and molecular structures in advanced materials research,[1] and polarized neutron scattering has the capability to investigate structural and magnetic properties through neutron spin polarization.[2,3] Enhanced neutron optics, capable of increasing the neutron flux at higher angles of reflection with thinner bilayer periods, enable novel types of optics.[4–7] A highly polarized neutron flux is preferable for any neutron scattering instrument, to decrease the measurement time and/or increase the statistics. The Fe/Si multilayer is the state-of-the-art materials system for neutron polarizer and analyzer optics used in neutron instruments.

Multilayer mirror optics are composed of two or more alternating materials repeated constructive interference, as described by Bragg's modified law. The reflected intensity is determined by the scattering length density (SLD) contrast, number of periods (N), and interface smoothness. Ideally, sharp interfaces yield optimal reflectivity. For multilayers with shorter period thicknesses designed to achieve high-angle neutron reflection, maintaining the interface quality becomes increasingly important. In this context, the Fe/Si multilayer system faces notable challenges, primarily due to the interface width between the Fe and Si layers. During growth a solid-state reaction between Fe and Si induces the formation of iron silicides at the interfaces, leading to broader interface widths of 8-10 Å.[8] This broadening reduces the reflectivity of the system because of a non-smooth transition in SLD contrast between the layers. While it is possible to compensate for these effects to some extent by increasing the number of periods, the magnetically dead silicide region negatively affects the polarization and can not be compensated for through this approach. In neutron optics, the m-value represents the range of reflectivity for neutron supermirrors, indicating the ratio of the maximum reflection angle of the supermirror to the critical angle of total reflection for a single-layer mirror of Ni. Higher m-values signify a broader range of reflectivity, allowing the supermirror to reflect neutrons at larger angles. A reduced interface width has been achieved using ion beam sputtering (IBS), but only to fabricate supermirrors up to *m* = 3.9.[4] In contrast, state-of-the-art Fe/Si multilayers deposited by DC magnetron sputtering can produce supermirrors up to *m* = 5.5.[9] Further, ion beam sputtering is a considerably more expensive method to produce equivalent optics. Reactive sputtering of Si with $N_2$ has also been assessed, but requires careful optimization depending on the supermirror layer thickness geometries.[10]

In our previous work, we addressed these challenges by co-sputtering $^{11}B_4C$ into Fe and Si layers, effectively preventing silicide-nanocrystal formation.[11] This resulted in enhanced reflectivity and polarization, magnetically soft Fe layers, and reduced diffuse scattering, even for bilayer periods as thin as 15 Å.[11] While incorporation of $^{11}B_4C$ can solve many critical issues with Fe/Si multilayers[11], co-sputtering has significant challenges in terms of uniformity over a large area, particularly when applied on large scale industrial deposition systems. These systems in most cases require modifications to sputtering geometry or sputtering targets, making co-sputtering less feasible for large-scale manufacturing with current industrial systems. Therefore, a more feasible alternative solution is to add the $^{11}B_4C$ as interlayers or barrier layers. Natural $B_4C$ interlayers have previously been used for improved X-ray optics.[12] The choice of $^{11}B_4C$ is based on its ability to form an amorphous alloy at the interfaces due to the B-metal bonding, in contrast to the amorphization of entire layers observed in co-sputtering. Further, if prevention of iron-silicide is successful, the SLD profile will avoid the almost nanometer-sized intermediary nuclear and magnetic step. Unlike co-sputtering, which dilutes Fe and Si and reduces SLD contrast, the use of $^{11}B_4C$ interlayers preserves high SLD contrast and should maintain a high reflectivity.

Here, we investigate the effect of varying $^{11}B_4C$ interlayer thicknesses on the performance of Fe/Si multilayers. Since the interlayer itself is part of the interface width, the interlayer thickness should not exceed the interface width, which for a pure Fe/Si multilayer can be up to 8.6 Å.[5] The minimum interlayer thickness is constrained by the size of a B atom to about 1 Å. To account for this, we have chosen to investigate interlayer thicknesses of 2 Å for thicker periods (100 Å), 1 Å for the thinnest periods (15 Å), and both 1 and 2 Å for intermediate period thicknesses (30 Å). Our analyses focus on reflectivity, polarization, and the impact on magnetization and coercivity from the addition of interlayers. This approach offers a practical comparison between traditional Fe/Si multilayers and those utilizing interlayers, particularly for industrial sputtering systems lacking co-sputtering capabilities. The findings aim to show that interlayer-based solutions are an accessible and effective alternative to co-sputtering.

## 2. Experimental details

Ion-assisted magnetron sputter deposition in a high vacuum environment (approximately $5.6 \cdot 10^{-5}$ Pa or $4.2 \cdot 10^{-7}$ Torr) was utilized to deposit Fe/Si and Fe/$^{11}B_4C$/Si/$^{11}B_4C$ multilayer thin films. The deposition system is described in detail elsewhere.[13] The multilayers were grown onto 001-oriented single-crystalline Si substrates measuring $10 \times 10 \times 1$ mm$^3$ in size, with a native surface oxide layer. Sputtering targets were Fe (99.95% pure, 75 mm diameter), Si (99.95% pure, 75 mm diameter), and 99.8% chemically pure, >90% isotopic purity $^{11}B_4C$ target with 50 mm diameter. Continuous operation of magnetrons during deposition, along with computer-controlled shutters for each target material, allowed precise control of atom fluxes, enabling multilayer deposition. The deposition rates for Fe and Si were both around 0.4 Å/s, while the rate for $^{11}B_4C$ was 0.08 Å/s. The $^{11}B_4C$ rate was calculated from the period thicknesses of multilayers with and without interlayers, as determined from X-ray reflectivity fitting. The substrates were kept at ambient temperature during deposition and rotated at 8 rpm to ensure even thickness distribution. For the first 3 Å of each layer, the substrate was kept at a floating potential followed by a -30 V substrate bias for the remaining thickness. The plasma was condensed towards the substrate by a magnetic field aligned with the substrate normal by a coil, enhancing the Ar-ion flux to the growing film.[13] The design parameters for all the deposited films are listed in Table 1.

X-ray reflectivity experiments were performed using a Malvern Panalytical Empyrean diffractometer equipped with Cu-$K\alpha$ radiation and a PIXcel detector. A Göbel mirror, along with a 0.5° divergence slit, was implemented on the incident beam side, while a parallel beam collimator and a 0.27° collimator slit were utilized on the diffracted beam side. Reflectivity data from X-rays were analyzed to calculate the multilayer period, the thickness of individual layers, and interface roughness, using the GenX3 software.[14]

Additionally, X-ray diffraction measurements were carried out with a Panalytical X'Pert diffractometer, using Bragg-Brentano geometry, scanning over a $2\theta$ range from 20° to 90°. For enhanced clarity, the analysis primarily focused on the 39.5° to 52° region.

Table 1. Summary of parameters for all multilayers investigated in this study. The listed layer thicknesses represent nominal values.

| Multilayers | $^{11}B_4C$ Layer Thickness [Å] | Period thickness ($\Lambda$) [Å] | Fe layer thickness/Si layer thickness [Å] | Number of periods ($N$) |
|---|---|---|---|---|
| *Multilayers with varying $^{11}B_4C$ inter layer thicknesses, periods, and total number of periods* | 0 and 2 | 100 | 50/50, 48/48 | 20 |
| | 0 and 1 and 2 | 30 | 15/15, 14/14, 13/13 | 20 |
| | 0 and 1 | 15 | 7.5/7.5, 6.5/6.5 | 20 |
| | 0 and 1 | 15 | 7.5/7.5, 6.5/6.5 | 40 |
| | 0 and 1 | 15 | 7.5/7.5, 6.5/6.5 | 80 |
| | 0 and 1 | 15 | 11.25/3.75, 10.25/2.75 | 20 |

The magnetic properties of the samples were assessed at room temperature using vibrating sample magnetometry (VSM) in a longitudinal setup, with measurements performed across a magnetic field range of -25 mT to 25 mT. Polarized neutron reflectometry (PNR) experiments were performed using the Morpheus instrument at the Swiss Spallation and Neutron Source (SINQ) located at the Paul Scherrer Institute (PSI) in Switzerland. PNR is sensitive to the spin-dependent scattering length density (SLD) of the sample, thus providing insights into the magnetization profile. The two spin states produce distinct reflectivity curves, with Bragg peaks occurring due to constructive interference. In these experiments, a polarized beam of neutrons was directed at small incidence angles ($\theta$) towards the sample, reflecting off of the sample before detection by a He-3 detector. Measurements were conducted with the samples in an external magnetic field of approximately 20 mT, covering a $2\theta$ range of 0° to 15°, using neutrons with a wavelength of 4.825 Å.

Transmission electron microscopy (TEM) cross-sectional samples were prepared using conventional mechanical polishing followed by argon ion etching at 5 keV, with a final etching step at 2 keV to remove surface damage. High-angle annular dark-field scanning transmission electron microscopy (HAADF STEM) at atomic resolution was performed using a double Cs-corrected Titan$^3$ 60-300 microscope in Linköping, operated at 300 keV. TEM images and selected area electron diffraction (SAED) patterns were acquired using a FEI Tecnai G2 microscope operated at 200 keV.

Ion beam analysis was carried out at the Tandem Laboratory at Uppsala University.[15] Time-of-Flight Elastic Recoil Detection Analysis (ToF-ERDA) measurements was performed at the 5 MV Pelletron accelerator utilizing with a $36$ MeV $I^{8+}$ beam at an incident angle of 67.5° and a recoil detection angle of 45.0°. The energy of the recoiled particles was measured using a gas ionization chamber, while C foils were employed to record the time-of-flight. Nuclear Resonance Analysis (NRA) was performed using the 350 keV ion implanter, employing the $^{11}B(p,\alpha)^8Be$ resonance at 163 keV to investigate the amount of $^{11}B$ in the samples.[16] All NRA measurements were performed at an incident angle of 60°. Additionally, Time-of-Flight Medium Energy Ion Scattering (ToF-MEIS) measurements were carried out with a 120 keV $He^+$ beam, recorded at a scattering angle of 160° and an incidence angle of 0°.

## 3. Results

### 3.1. X-ray Reflectivity, XRR

X-ray reflectivity measurements and simulations using the GenX software were made for all multilayers investigated in this work. Figure 1(a) compares the XRR profiles of multilayers a period thickness of 100 Å, both without and with 2 Å $^{11}B_4C$ interlayers. The Fe/Si multilayer without interlayers exhibits 10 Bragg reflections while the multilayer with interlayers show up to 13 sharp Bragg peaks, extending to $2\theta = 10°$. It can be observed that the odd-numbered Bragg peaks are stronger for the interlayered sample. The inset in Figure 1(a) shows the fitted SLD depth profiles, revealing that the Si-on-Fe interface increases slightly from 5 Å to 6 Å for the interlayered multilayer, while the Fe-on-Si interface width decreases significantly 13.3 Å to 7.6 Å. This implies and can be visually determined from the SLD profiles that the interlayered sample has a more homogeneous/symmetrical profile.

Figure 1(b) shows the XRR comparison of the multilayers with a nominal period thickness of $\Lambda = 30$ Å, with the inset showing the first Bragg peak on a linear scale. Higher intensities for the 1 Å and 2 Å interlayered multilayers is evident, even after considering the scattering attenuation due to the Debye-Waller factor at higher Bragg angles. Simulations confirm similar interface widths of $\sigma = 6.0 - 6.2$ Å for Si-on-Fe interface widths across all multilayers, consistent with results for $\Lambda = 100$ Å multilayers in Figure 1 (a), where interlayers have minor effect in reducing the interface width when deposited on top of the Fe layers. However, the Fe-on-Si interface width decreased from 8.2 Å (no interlayer) to 6.1 Å (1 Å interlayer) to 5.2 Å (2 Å interlayer). Since the Si-on-Fe interface width slightly increases with the 2 Å interlayers, the optimum interlayer thickness is considered to be between 1 Å and 2 Å.

For $\Lambda = 15$ Å multilayers, where the individual layer thicknesses is around 7.5 Å i.e., comparable to the expected interface widths of pure Fe/Si multilayers, multilayers without interlayers and with 1 Å interlayers were investigated. Different number of periods (N = 20, 40, 80), to enhance the reflectance from these thin periods, were studied to evaluate reflectance and accumulated roughness. Figure 1(c) shows the XRR profiles for these multilayers, vertically shifted for clarity. The first-order Bragg peaks for interlayered samples are shifted to higher angles compared to the pure Fe/Si samples, indicating slightly thinner periods than expected. This further proves that even though intensity decreases with scattering angle, the intensity in our comparison is still larger when $^{11}B_4C$ layers are incorporated than for the Fe/Si case.

Additionally, the pronounced Kiessig fringes observed for interlayered samples highlight smoother and more well-defined interfaces compared to the pure Fe/Si samples. Simulations for N = 80 multilayers confirm these observations, yielding average interface widths of 9.5 Å and 5.2 Å, for the Fe/Si and interlayered samples, respectively. Since the interface width of at least one of the interfaces for the Fe/Si sample often exceeds the nominal layer thickness, it can be concluded that no pure Si layers exist in these samples, instead the entire non-magnetic layer is intermixed with Fe. Therefore, interface widths are reported as averages between the Si-on-Fe and Fe-on-Si interface widths.

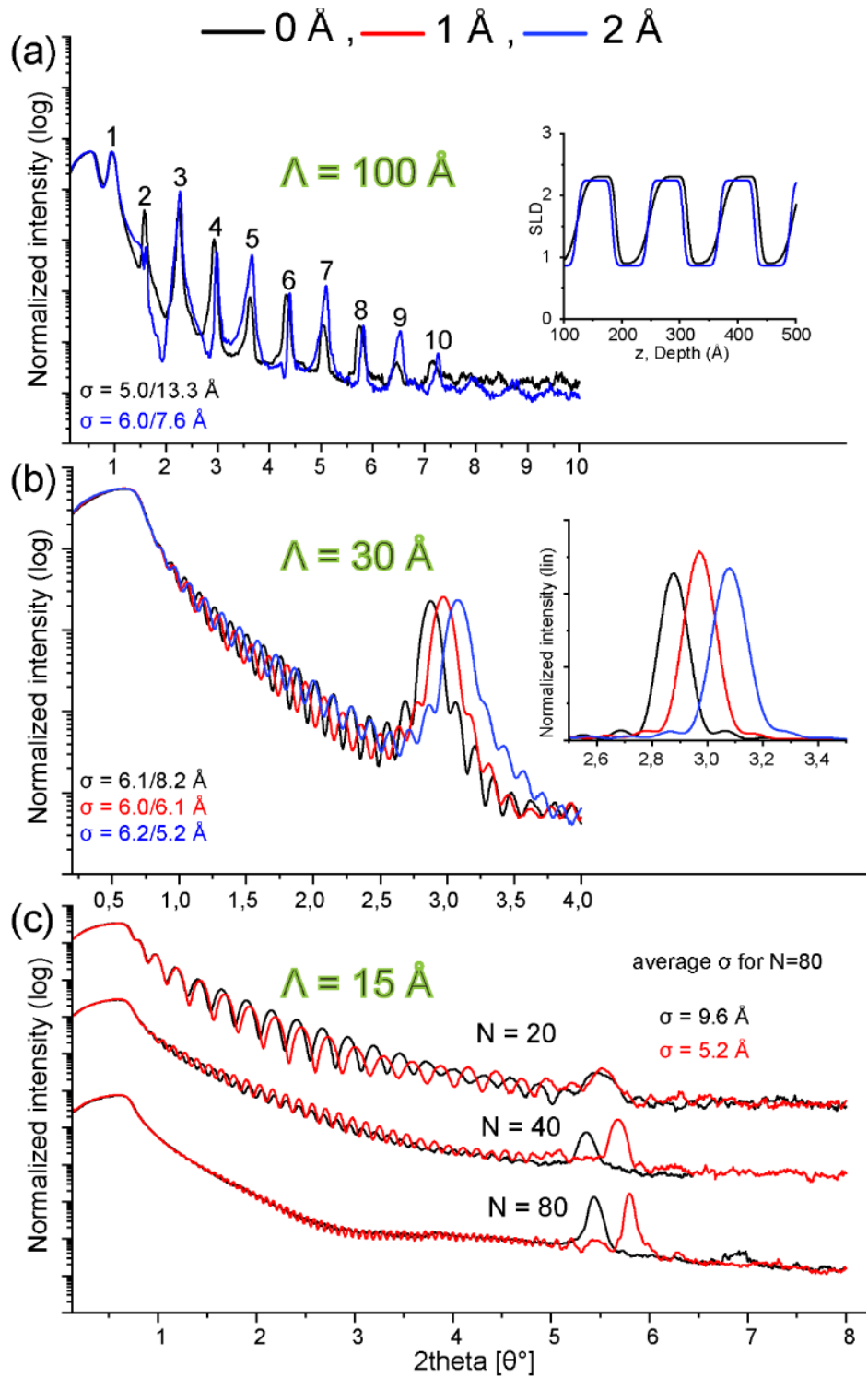

Figure 1: X-ray reflectivity (XRR) data normalized to their critical edges, along with simulated interface widths, $\sigma$, for Si-on-Fe/Fe-on-Si interfaces, for multilayers without $^{11}B_4C$ interlayers (black), with 1 Å $^{11}B_4C$ interlayers (red), and with 2Å $^{11}B_4C$ interlayers (blue) for: (a) $\Lambda =100$ Å multilayers with $N = 20$. The inset shows the corresponding SLD depth profiles. (b) $\Lambda = 30$ Å multilayers with $N = 20$. The inset shows the first Bragg peak region on a linear scale. (c) $\Lambda = 15$ Å multilayers with $N = 20, 40,$ and $80$, vertically shifted for clarity.

## 3.2. STEM/TEM and SAED

To further characterize the distribution and effect of $^{11}B_4C$ interlayers at the interfaces large period multilayers with a period of $\Lambda = 100$ Å (N = 20) were analyzed using analytical TEM. The analysis of cross-sectional samples from two multilayers - Fe/Si and Fe/Si with 2 Å $^{11}B_4C$ interlayers - is shown in Figure 2. High-resolution TEM comparison in Figure 2(a) and (b) reveal a more distinct Fe-rich (dark) and Si-rich (bright) layer structure in the pure Fe/Si multilayer, while the interlayered multilayers show diffuse interfaces with thinner brighter regions. In the pure Fe/Si multilayer, the Fe-rich layers are more textured, in contrast to the interlayered structure, where the Fe-rich layers exhibit less texture. The Si-rich layers in both multilayers appear amorphous. Selected area electron diffraction (SAED) patterns confirm the texture of the Fe-rich layers in the Fe/Si sample, with no clear distinction between Fe and Fe-silicide reflections, likely due to overlap of the two body-centered cubic (bcc) phases. This observation is consistent with X-ray diffraction (XRD) results (see Supplementary Information). Both XRD and SAED data suggests the presence of superlattice fringes, particularly pronounced in Fe/Si multilayer, around the 110 reflections of Fe and Fe-silicide. This supports the crystalline nature of the Fe-rich layers. HAADF-STEM images confirm the TEM findings, showing more diffuse interfaces in the interlayered sample compared to the pure Fe/Si multilayer. However, the Fe-containing layers, which appear bright in HAADF-STEM, exhibit a bright core and relatively flat interfaces. Energy-dispersive X-ray spectroscopy (EDX) (not shown) and electron energy loss spectroscopy (EELS) elemental maps further validate these observations., revealing that, despite the diffuse interfaces in STEM images, the Fe and Si interfaces are abrupt in the interlayered multilayer. EELS elemental maps, presented in Figures 2(g) and 2(h) confirm the Fe core within the Fe-rich layers and demonstrate reduced Si diffusion into Fe for the interlayered sample compared to the pure Fe/Si multilayers. EDX analysis, consistent with the EELS results, revealed Fe, Si, and Ar compositional modulations in both samples, which are therefore not presented here.

An additional observation from the EDX and EELS maps reveals an unexpected compositional modulation of argon, the sputtering gas used during deposition. Argon atoms appear to be trapped within the Si layers, with a higher concentration localized toward the center of the Si layers in the interlayered sample. This may be attributed to Si self-confinement resulting from reduced diffusion. This finding is reported here for the first time, and further investigation will be conducted in future studies, as it could potentially influence the scattering length density (SLD) contrast in mirror applications.

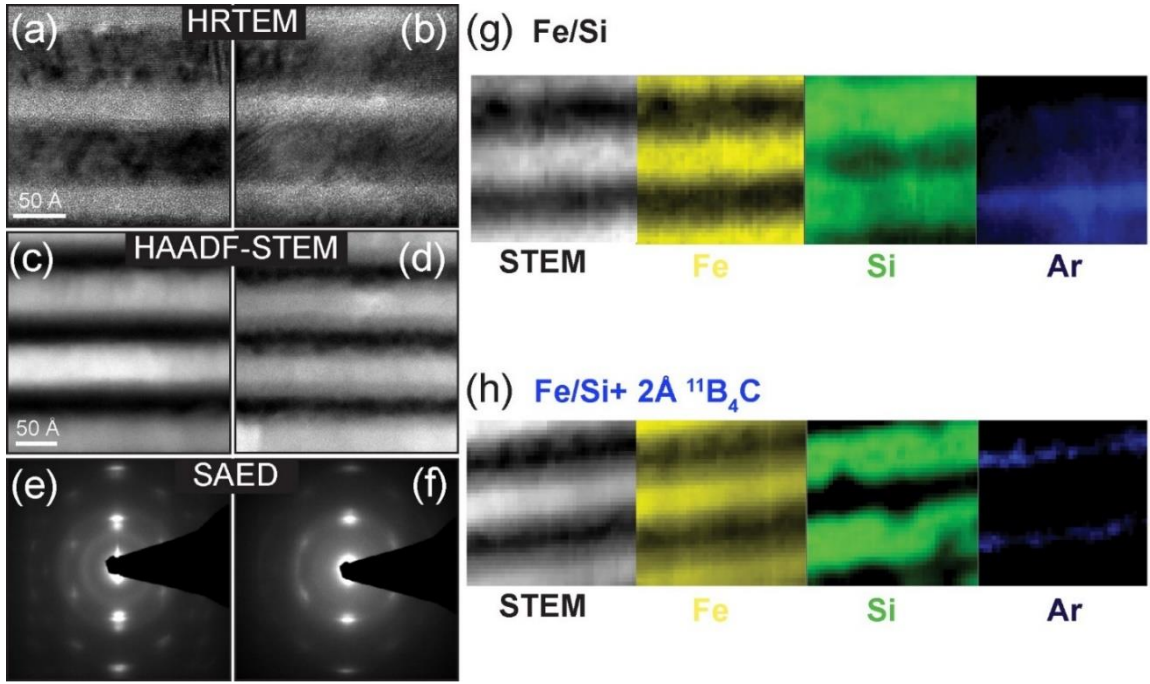

*Figure 2: TEM/STEM characterization of Λ = 100 Å Fe/Si and Fe/Si + 2 Å $^{11}B_4C$ interlayers, respectively: HR-TEM images (a and b), HAADF-STEM images (c and d), SAED patterns of the films obtained along the <011> zone axis of the Si(001) substrate (e and f), EELS elemental maps of Fe, Si, and Ar and corresponding HAADF-STEM images (g and h).*

No signals for the lighter elements, boron and carbon, were detected in the EELS and EDX maps, likely due to their low concentration. However, EELS analysis indicates that the interlayers reduce Si diffusion into the Fe layers, resulting in relatively sharper and more abrupt interfaces.

### 3.3. Ion Beam Analysis, IBA

XRR analysis revealed that the introduction of interlayers effectively reduces interface width in the multilayers. However, analytical TEM did not provide any direct evidence of the presence of $^{11}B_4C$ layers or their role in achieving smooth or abrupt interfaces. On the contrary, interfaces appear rougher in the TEM images, but with a sharper contrast between Fe-rich and Si-rich layers, suggesting that $^{11}B_4C$ hinders the intermixing to some extent. To investigate the presence of $^{11}B_4C$ and further study the silicide formation, we investigated multilayers with a period of Λ = 100 Å (N = 20) using multiple ion beam analysis methods. As shown in Figure 3 (a-b), ToF-ERDA determined the average film composition, revealing that 2 at.% of the Fe/Si + 2 Å $^{11}B_4C$ multilayer consisted of $^{11}B$ + C. Due to limited depth resolution, individual layers could not be resolved in the profiles. Additionally, about 2 at.% of Ar was detected in the films, supporting the TEM results and indicating that Ar is trapped during the deposition process rather than being introduced during the TEM sample preparation.

A clear advantage of $^{11}B_4C$ interlayers over co-sputtered $^{11}B_4C$ multilayers is the lower amount of $^{11}B_4C$, which affects the SLD contrast for reflectivity and polarization performance. For the Λ = 100 Å multilayer period, the 2 Å thick $^{11}B_4C$ interlayers correspond to $^{11}B$ + C is 2 at.% of the multilayer. In comparison, for Λ = 30 Å, the same $^{11}B_4C$ interlayers correspond to ~7 at.% of the multilayers, about half the 15 at.% of $^{11}B$ + C observed for co-sputtered Fe/Si + $^{11}B_4C$ multilayers in our previous work.

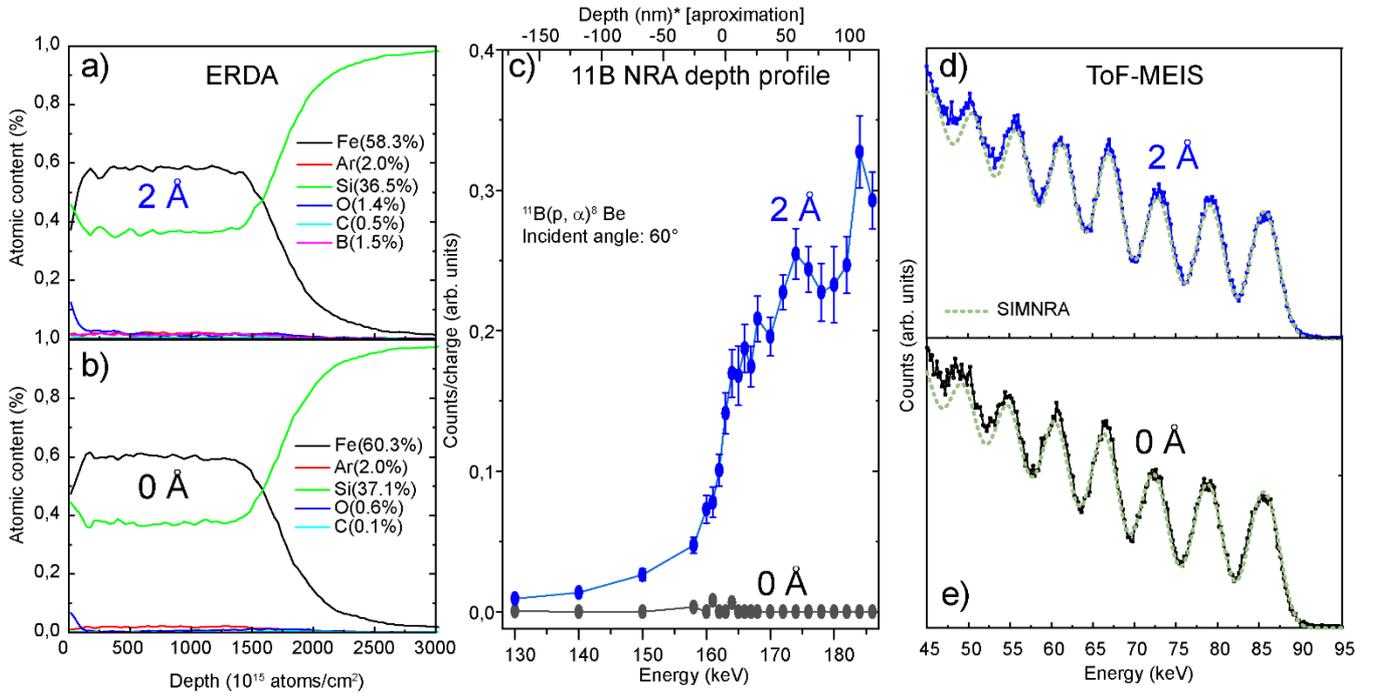

*Figure 3: Ion beam analysis of of Λ = 100 Å Fe/Si and Fe/Si + 2 Å $^{11}B_4C$ interlayers multilayers, respectively. ERDA analysis (a and b), $^{11}B$ NRA depth profile (c), Tof-MEIS analysis and simulations.*

Because of the sensitivity and selectivity to $^{11}B$ of the Nuclear Reaction Analysis (NRA), we performed a $^{11}B$ depth profile to investigate the boron composition modulation in the multilayers, as shown in Figure 3(c). Using the 163 keV energy of resonance (width = 6 keV), and considering the stopping power of protons in Si and Fe at 170 keV with an incident angle of 60°, the depth resolution was estimated to be in the range 13-30 nm. Although boron was detected, the limited depth resolution prevented the identification of clear depth features.

In addition to boron detection, we also attempted to investigate potential differences in silicide thickness between samples with and without interlayers using TOF-MEIS.[16,17] The measurements were performed at a 0° incidence angle with a depth resolution of approximately 7 nm for areal density determination, and at a 70° incidence angle, providing higher depth resolution of roughly 1.7 nm, however above the expected silicide layers. ToF-MEIS measurements revealed consistent layer thicknesses across the ~2 mm² measurement area. Since ToF-MEIS is sensitive to differences in the thickness of single layers, we can conclude that the consistent layer thickness also applies to the topmost layer. The areal densities of Fe and Si from ToF-MEIS are presented in Table 2. For Si, the uncertainty in areal density is influenced by the stopping cross section uncertainties of both Si and Fe, the scattering potential, statistical variations from total counts, and the nominal $^{11}B_4C$ values from deposition. In total the Si areal density uncertainty is estimated to 5% for the areal density for the Fe/Si sample and 7% for the Fe/Si + 2 Å $^{11}B_4C$ sample. For Fe areal densities uncertainties are lower, about 4%, since the signal of He ions scattered on the first Fe layer is directly visible in the spectrum. The main sources of uncertainties are the scattering potential and the measurement statistics. The measurements indicate a decrease in Fe areal density with increasing $^{11}B_4C$, while the Si areal density remains unchanged within the error limits. However, due to limited depth resolution, it was not possible to estimate the silicide thickness for either samples.

*Table 2: Areal densities based on ToF-MEIS experiments performed with 120 keV He+, a scattering angle of 160° and an incident angle of 0° degree. The areal densities of Si and Fe were extracted using SIMNRA simulations. The areal*

density of $^{11}B_4C$ was assumed based on the nominal values as an input parameter for the simulations. All values are in units of $10^{15}$ atoms/cm$^2$.

|  | Fe/Si | Fe/Si + 2Å $^{11}B_4C$ |
|---|---|---|
| Si | 39.0 ± 2.0 | 38.5 ± 2.7 |
| Fe | 46.5 ± 1.9 | 41.1 ± 1.6 |
| $^{11}B_4C$ | 0 | 2.6 ± 0.2 |

## 3.4. Vibrating sample magnetometry, VSM

Figure 4(a) presents the magnetization as a function of the external magnetic field for Fe/Si multilayers with a thickness of Λ = 100 Å (N = 20). Data are presented for both pure Fe/Si (black) and Fe/Si with 2 Å $^{11}B_4C$ interlayers (blue). As expected, the magnetization amplitude for the interlayered sample is slightly reduced, due to the lower overall Fe content compared to the pure Fe/Si multilayer. However, the coercivity curve of the interlayered sample deviates from the typical ferromagnetic response. Specifically, the presence of additional features, resembling shoulders, suggests that certain magnetic domains or layers exhibit increased resistance to magnetization before reaching saturation. While the origin of this enhanced resistance, whether due to magnetic domain behavior or interlayer coupling, can not be conclusively determined from the current data, it does not affect the main objective of the study or its potential applications. Notably, the saturation field remains similar for both samples, approximately 5 mT. The saturated magnetization values are 0.0177 emu for the Fe/Si multilayer and 0.0170 emu for the Fe/Si + 2 Å $^{11}B_4C$ sample.

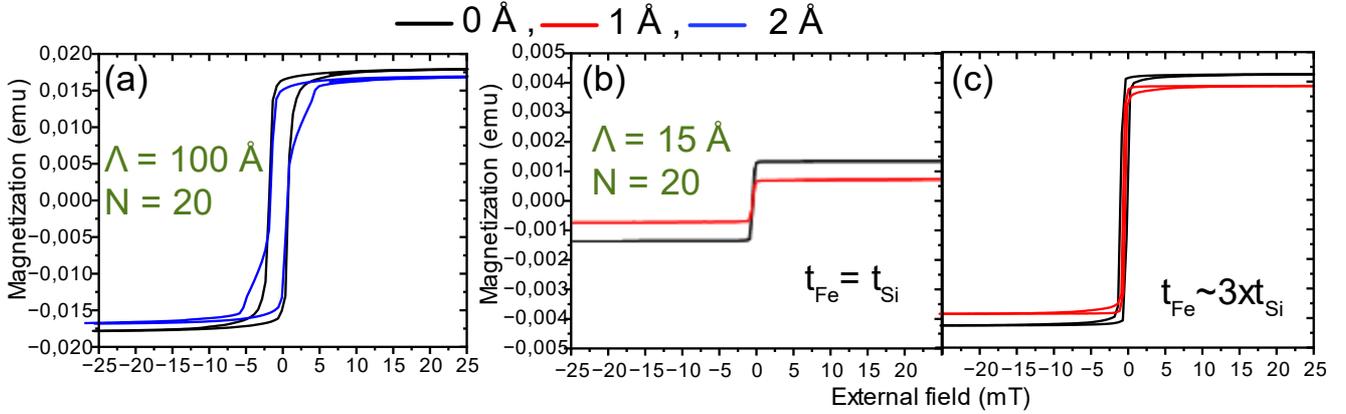

Figure 4. Vibrating sample magnetometry (VSM) of Fe/Si and Fe/Si + 2 Å $^{11}B_4C$ interlayers, (a) Λ = 100 Å, (b) of Λ = 15 Å and (c) also with Λ = 15 Å but with a Fe thickness being 3 times thicker than the Si thickness. All samples have N = 20 periods.

Figure 4(b) presents the VSM results for samples with a period thickness of Λ = 15 Å and N = 20, incorporating 0 and 1 Å interlayers. The observed low coercivity is typical for these thin, amorphous multilayers, where a significant portion of the bilayer consists of silicide. The magnetization values for Fe/Si and Fe/Si + 1 Å $^{11}B_4C$ are 0.0013 and 0.0007 emu, respectively. When considering the total volume of the Fe layers, the Fe atom count in the Λ = 15 Å samples is approximately 6.67 times lower for Fe/Si and 7.38 times lower for Fe/Si + 1 Å $^{11}B_4C$ compared to the Λ = 100 Å samples shown in Figure 4(a). Scaling the magnetization values from Figure 4(a) (0.0177 emu for Fe/Si and 0.0170 emu for Fe/Si + 1 Å $^{11}B_4C$) by these factors predict magnetization values of 0.0026 and 0.0022 emu, respectively. These predictions are significantly higher than the experimentally measured values, which can be attributed to the large silicide fraction in the $\Lambda$ = 15 Å multilayers. When the Fe layer thickness was increased by a factor of three relative to Si in the 15 Å period multilayers, the corresponding VSM data, shown in Figure 4(c), indicate saturated magnetizations of approximately 0.0042 emu for Fe/Si and 0.0039 emu for Fe/Si + 1 Å $^{11}B_4C$. These values are in close agreement with the expected values of 0.0040 and 0.0035 emu, based on the magnetization scaling in the 100 Å period multilayers. This confirms that the lower magnetization observed in Figure 4(b) is due to the large silicide fraction, which remains largely unaffected by the 1 Å interlayer.

## 3.5. Polarized neutron reflectivity, PNR

Figure 5(a-c) shows the spin-up PNR and corresponding fitting curves of Fe/Si multilayers with 0, 1, and 2 Å interlayers of $^{11}B_4C$, multilayers with $\Lambda$ = 30 Å and N = 20. Measurements were focused on the critical angle region and the first order Bragg peak due to the limited beam time. At the Bragg peak, the spin-up intensity for samples with 0, 1, and 2 Å interlayers were measured to 92, 167, and 179 counts, respectively. The 94% increase in reflectivity from 0 to 2 Å interlayers, indicates a significant reduction in interface width with the addition of $^{11}B_4C$ interlayers. The polarization at the Bragg peak position was calculated to be 53%, 79%, and 89% for 0, 1, and 2 Å interlayered samples, respectively, which highlight the effectiveness of $^{11}B_4C$ interlayers.

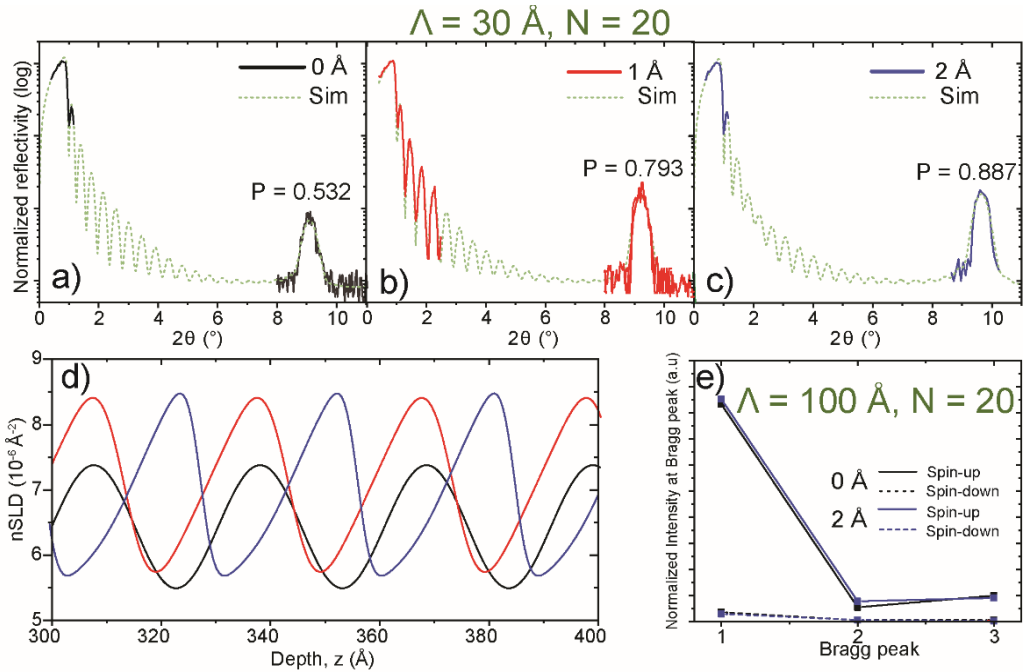

*Figure 5. Spin-up polarized neutron reflectivity (PNR) measurements of (a-c): Fe/Si multilayers with 0, 1, and 2 Å interlayers of 11B4C, respectively, where all multilayers had $\Lambda$ = 30 Å and N = 20. (d) shows the corresponding nuclear SLD profiles stemming from the fits in a-c. (e) shows the Bragg peak intensities of the first, second, and third order Bragg peaks for both spin-up (solid) and spin-down (dashed) neutron reflectivity for $\Lambda$ = 100 Å and N = 20. These multilayers are also used for the IBA and XRR analysis.*

To further illustrate the difference between the samples, Figure 5(d) shows the corresponding nuclear SLD depth profile from 300 to 400 Å above the substrate, based on the fits shown in Figure 5(a-c). The SLD depth profiles reveal that the interlayered samples have higher oscillation amplitudes. Figure 5(e) shows the reflectivity value for the first, second, and third Bragg peak order for spin-up and spin-down for the two samples Fe/Si and Fe/Si + 2 Å $^{11}B_4C$ where $\Lambda$ = 100 Å and N = 20. It shows that the Bragg peak intensities do not differ significantly in reflectivity and only show a slight improvement in polarization for all peaks for the $^{11}B_4C$ sample.

## 4. Discussion

The primary objective of this study is to investigate the effects of incorporating $^{11}B_4C$ interlayers into Fe/Si multilayers, with a focus on interface definition, silicide formation, reflectivity, polarization, and potential impacts on the magnetic properties of the multilayers. The motivation comes from the need to provide an alternative fabrication approach, compatible with existing industrial deposition techniques. This approach aims to create multilayers with properties comparable to those of the recently developed polarized neutron mirrors where $^{11}B_4C$ was integrated into Fe and Si layers through co-sputtering.[11]

The performance of neutron mirrors is closely related to the quality of their interfaces, which becomes increasingly critical as the bilayer period thickness decreases. Smooth interfaces enhance the ability to reflect and polarize neutrons at higher scattering angles, a key factor in achieving high performance. State-of-the-art Fe/Si supermirrors with a period thickness of 5.5 Å can achieve reflectivity up to up $q = 0.12$ Å$^{-2}$.[9] To assess the potential for high-angle reflectivity in multilayers containing $^{11}B_4C$ interlayers, the bilayer period thickness must be sufficiently thin. In this study, Fe/Si multilayers with 1 Å and 2 Å $^{11}B_4C$ interlayers were deposited with period thicknesses of 15 Å, 30 Å, and 100 Å. These were compared with pure Fe/Si multilayers to evaluate their mirror performance.

### 4.1. Reflectivity and polarization enhancement

X-ray reflectivity (XRR) analysis revealed enhanced reflectivity for all multilayers containing $^{11}B_4C$ interlayers, including those with thinner periods, such as the 15 Å multilayers with individual layer thicknesses of approximately 7 Å. Based on the analysis of interface widths from the reflectivity data, the optimal interlayer thickness was found to be 2 Å for multilayers with a period of $\Lambda = 100$ Å, and 1 Å for those with a period of $\Lambda = 15$ Å. The observed increase in reflectivity with the number of periods (N = 20, 40, and 80) for the thinner-period multilayers ($\Lambda = 15$ Å) with 1 Å $^{11}B_4C$ interlayers demonstrates that the reflectivity gain scales with the number of periods. This also suggests that introducing $^{11}B_4C$ interlayers does not lead to accumulated roughness, a key requirement for high-performance supermirror fabrication. The improvement in XRR for the $\Lambda = 15$ Å multilayers with $^{11}B_4C$ interlayers is comparable to that observed in $^{11}B_4C$ co-sputtered multilayers [see Supplemental Information].

Consistent with the XRR results, polarized neutron reflectivity (PNR) measurements for 30 Å period multilayers (Figure 5) showed a clear increase in both reflectivity and polarization upon the addition of 1 Å and 2 Å $^{11}B_4C$ interlayers. Specifically, the polarization at the Bragg peak positions increased by 66% for the 2 Å $^{11}B_4C$ interlayer ($\Lambda = 30$ Å), exceeded the improvements observed in $^{11}B_4C$ co-sputtered multilayers with similar periodicities. Combined XRR and PNR results indicate that the interlayer approach enables reflection and polarization at $q = 0.44$ Å$^{-1}$ (m = 16.5 for 15 Å period), which corresponds to a vector reflection approximately 3.5 times higher than that achieved by state-of-the-art polarized neutron optics.

While reflectivity and polarization improvements were more pronounced for thinner-period multilayers, the 100 Å multilayers exhibited more modest enhancements compared to the $^{11}B_4C$ co-sputtered multilayers from previous studies.[11] This suggests that the advantages of the $^{11}B_4C$ interlayers are more significant for thinner periods, where the interface width and magnetic variations play a more significant role. Therefore, the inclusion of interlayers has a more substantial impact on the performance of thinner-period multilayers.

## 4.2. Interface width, amorphization and silicide formation

Although the increased reflectivity and polarization observed with the introduction of $^{11}B_4C$ interlayers is clear, the underlying mechanisms responsible for these improvements remain a critical focus of investigation. We hypothesize that the $^{11}B_4C$ interlayer functions primarily as a barrier, inhibiting the bonding and/or mixing of Fe with Si, thereby preventing the formation of iron-silicide phases or heavily mixed regions at the interfaces. This would contribute to the enhancement of interface quality and performance. Additionally, the strong affinity of B for metals, particularly iron, contributes to amorphization of the Fe layers at the interfaces $^{11}B_4C$ is known to amorphize metals, as boron has a strong affinity for binding with metal atoms, particularly iron, and in doing so, it may hinder the formation of Fe-Fe bonds. In the context of X-ray optics, the amorphization of the metal layers at the interfaces is beneficial, as it can improve interface quality by reducing roughness typically caused by nanocrystallites or nanofacets. This reduction in interface roughness is crucial for optimizing the performance of multilayer mirrors and improving their reflectivity and polarization capabilities.

The Bragg reflections observed in XRR and PNR measurements, along with reflectivity fitting for multilayers with various periods, confirm a reduction in interface width for all interlayered samples, including those with the thinner 15 Å periods. Reflectivity intensity is influenced by the scattering length density (SLD) contrast between the layers and the interface width, but in this case, the reflectivity gain can not be attributed to changes in the SLD contrast between Fe and Si, as this remains unaffected by the interlayer. Instead, the improvement in reflectivity can be attributed solely to reduced interface widths, as confirmed by the fits. These indicates that, without an interlayer, the interface width between Fe and Si is significantly larger compared to the interlayered samples. SLD profiles (Figures 1(a) and 5(d)) show more abrupt SLD contrasts with $^{11}B_4C$ interlayers, supporting the reduction in interfacial roughness.

The effect of interlayers on interface width is found to be asymmetric for the two interfaces, Si-on-Fe and Fe-on-Si. Simulations revealed that only minor differences in the interface widths of Si-on-Fe interfaces with and without interlayers, while a significant reduction in the Fe-on-Si interface width was observed. For example, for a period of $\Lambda = 30$ Å, the Fe-on-Si interface width decreased from 8.2 Å to 6.1 Å with a 1 Å interlayer and further to 5.2 Å with a 2 Å interlayer. The larger interface width for Fe-on-Si interfaces in the absence of interlayers is attributed to enhanced diffusion and intermixing, driven by a combination of Fe-backscattered neutrals, surface energies, and phase-formation thermodynamics during deposition. Studies by Romano et al.[18] have shown that asymmetric magnetic and nonmagnetic silicide layers form at interfaces, with their thickness depending on composition and deposition conditions. In our study, ion-assisted magnetron sputter deposition, combined with $^{11}B_4C$ interlayers, affects surface energies and diffusion processes, influencing silicide formation. This appears to be particularly pronounced at the Fe-on-Si interface, suggesting that the interlayers play a critical role in reducing interfacial roughness by altering the diffusion dynamics and silicide formation mechanisms.[19]

While analytical TEM could not directly resolve interface asymmetries, STEM/EELS and EDX analyses revealed that interlayers reduced Si diffusion into Fe layers, resulting in relatively sharp interfaces. No boron or carbon signals were detected in these analyses, but ERDA confirmed the presence of $^{11}B_4C$, with a 2 Å interlayer in 15 Å period multilayers corresponds to approximately 7 at.% $^{11}B_4C$. This finding highlights an advantage of $^{11}B_4C$ interlayers over co-sputtered $^{11}B_4C$ multilayers, as the lower $^{11}B_4C$ content minimizes the scattering length density (SLD) contrast dilution, thereby preserving reflectivity and polarization performance.

Despite extensive ion beam analysis to determine silicide layer thicknesses, poor resolution limited definitive conclusions about interlayer effects. ToF-MEIS measurements, however, indicated consistent layer thicknesses across a ~2 mm$^2$ measurement area for all multilayers, reflecting the overall quality and high uniformity of the multilayers.

For thicker periods of 100 Å, XRR showed significant reduction in interface width by adding 2 Å interlayers, especially on the Fe-on-Si interface. STEM and EELS analysis suggested reduced Si diffusion into Fe layers, however, PNR measurements did not detect significant difference in polarization compared to pure Fe/Si multilayers. In comparison, $^{11}B_4C$ co-sputtering was shown to be advantageous for thinner as well as thicker periods. The key difference in the two designs is the amorphization of Fe and Fe silicide layers, which becomes prominent for thicker periods. For thicker periods, 2 Å interlayers reduce silicide formation and only amorphized the interface region, as seen in the STEM contrast images, but can not overcome the roughness generated by highly textured Fe and Fe silicide nanocrystallites, and the interface width remains high compared to thinner periods. In contrast, in $^{11}B_4C$ co-sputtered amorphous multilayers, the interface width does not depend on the multilayer period.

## 4.3. Effect on saturation magnetization and coercivity

When investigating an alternative design of $^{11}B_4C$ co-sputtered multilayers, it is essential to understand how the presence of interlayers affects magnetization and coercivity. For Fe/Si multilayers with a bilayer period of $\Lambda = 100$ Å, the saturation magnetization of the pure Fe/Si sample was measured at 0.0177 emu, while the interlayered sample with a 2 Å interlayer showed a value of 0.017 emu, corresponding to a 5.5% decrease, which is consistent with the reduced Fe content in the interlayered sample. The external field required for saturation is similar in both samples, around 5 mT. However, the hysteresis loop for the interlayered sample is less steep during magnetization reversal, though this does not impact practical performance, as polarizers and analyzers operate under a saturated external field. For thinner-period multilayers (15 Å), both with and without interlayers, the low magnetization suggests significant silicide formation. The 1 Å interlayer is insufficient to suppress this effect. To mitigate silicide formation and improve magnetic properties, either a 2 Å interlayer or increasing the Fe layer thickness would yield better results.

## 5. Conclusion

In conclusion, incorporating $^{11}B_4C$ interlayers into Fe/Si multilayers significantly enhances interface quality by reducing interface width and preventing excessive Si diffusion into Fe. This results in improved reflectivity and polarization, particularly for thinner-period multilayers, enabling high-angle neutron reflectivity beyond the capabilities of conventional Fe/Si supermirrors. The optimal interlayer thicknesses of 1 Å for 15 Å period and 2 Å for 100 Å period multilayers improve interface sharpness and prevent iron-silicide formation, which would otherwise degrade both magnetic properties and neutron optical performance. However, characterizing the elemental distribution at the interfaces remains challenging due to the limitations of current analytical techniques.

While the interlayer approach yields significant improvements in reflectivity and polarization for thinner-period multilayers, the impact is less pronounced for thicker periods (100 Å) due to roughness caused by Fe and silicide nanocrystallites. The inclusion of interlayers also results in a slight decrease in saturation magnetization, consistent with the reduced Fe content. The introduction of $^{11}B_4C$ interlayers, however, does not affect coercivity, which limits their potential for applications requiring low coercivity for polarized neutron mirrors. Overall, $^{11}B_4C$ interlayers offer a promising strategy for enhancing the performance of Fe/Si multilayers, particularly for polarized neutron optics and supermirror designs, though further advancements in interface characterization are needed to fully understand their impact.

# Acknowledgments

The authors gratefully acknowledge funding from the Swedish Government Strategic Research Area in Materials Science on Functional Materials at Linköping University (Faculty Grant SFO-Mat-LiU No. 2009 00971). Financial support was also provided by the Swedish Research Council (VR) through project grants 2019-04837 (F.E.), 2018-05190 (N.G.), and 2021-03826 (P.E.), as well as by the Hans Werthén Foundation (grant 2022-D-03 to A.Z.), the Royal Academy of Sciences Physics Grant PH2022-0029 (A.Z.), the Lars Hiertas Minne Foundation (grant FO2022-0273 to A.Z.), the Längmannska Kulturfonden (grant BA23-1664 to A.Z.), and the SNSS travel grant (M.F.). Additional support was provided by the Knut and Alice Wallenberg Foundation through the Wallenberg Academy Fellows program (KAW-2020.0196 to P.E.). This work was partly conducted on the Morpheus neutron reflectometer at the SINQ spallation source, Paul Scherrer Institute, Switzerland. We thank Materials Characterisation Laboratory at ISIS for their support. We also thank ARTEMI, the Swedish National Infrastructure for Advanced Electron Microscopy, for access to the TEM and assistance with TEM analysis, and Ingemar Persson for help analyzing EELS data.